\documentstyle[preprint,epsfig,aps]{revtex} 
\draft
\begin{document}\tighten
\title{Brane cosmological  perturbations}
\author{David Langlois}
\address{D\'epartement d'Astrophysique Relativiste et de Cosmologie (UPR 176),\\
Centre National de la Recherche Scientifique,\\
Observatoire de Paris, 92195 Meudon Cedex, France}
\date{\today} 
\maketitle

\def\beq{\begin{equation}}
\def\eeq{\end{equation}}
\def\K52{{\kappa^2}}
\def\C{{\cal C}}
\def\lamb{{\rho_{\Lambda}}}

\newcommand{\A}{A}
\newcommand{\B}{B}
\newcommand{\mmu}{\mu}
\newcommand{\mnu}{\nu}
\newcommand{\ii}{i}
\newcommand{\jj}{j}
\newcommand{\jl}{[}
\newcommand{\jr}{]}
\newcommand{\ml}{\sharp}
\newcommand{\mr}{\sharp}

\newcommand{\da}{\dot{a}}
\newcommand{\db}{\dot{b}}
\newcommand{\dn}{\dot{n}}
\newcommand{\dda}{\ddot{a}}
\newcommand{\ddb}{\ddot{b}}
\newcommand{\ddn}{\ddot{n}}
\newcommand{\pa}{a^{\prime}}
\newcommand{\pb}{b^{\prime}}
\newcommand{\pn}{n^{\prime}}
\newcommand{\ppa}{a^{\prime \prime}}
\newcommand{\ppb}{b^{\prime \prime}}
\newcommand{\ppn}{n^{\prime \prime}}
\newcommand{\fda}{\frac{\da}{a}}
\newcommand{\fdb}{\frac{\db}{b}}
\newcommand{\fdn}{\frac{\dn}{n}}
\newcommand{\fdda}{\frac{\dda}{a}}
\newcommand{\fddb}{\frac{\ddb}{b}}
\newcommand{\fddn}{\frac{\ddn}{n}}
\newcommand{\fpa}{\frac{\pa}{a}}
\newcommand{\fpb}{\frac{\pb}{b}}
\newcommand{\fpn}{\frac{\pn}{n}}
\newcommand{\fppa}{\frac{\ppa}{a}}
\newcommand{\fppb}{\frac{\ppb}{b}}
\newcommand{\fppn}{\frac{\ppn}{n}}

\def\d{{\nabla}}
\def\D{{\tilde\nabla}}
\def\hh{{\hat h}}
\def\dd{{\delta}}
\def\g{{\tilde\gamma}}
\def\s{{\gamma}}
\def\vB{{\bar B}}
\def\vxi{{\bar \xi}}

\par\bigskip

\begin{abstract}
We address the question of cosmological perturbations in the context of 
brane cosmology, where our Universe is a three-brane where matter is confined,
whereas gravity lives in a higher dimensional spacetime. 
The equations governing the bulk perturbations are computed in the 
 case of a general warped universe. The results are then specialized 
to the case of a five-dimensional spacetime, scenario which has recently 
attracted a lot of attention. In this context, we decompose the perturbations
into `scalar', `vector' and `tensor' modes, which are familiar in the 
standard theory of cosmological perturbations. 
The junction conditions, which relate the metric perturbations to the 
matter perturbations in the brane,  are then computed.
\end{abstract}

\section{Introduction}
Recently has emerged the fascinating idea that extra dimensions (beyond our 
familiar time and three spatial dimensions) could be large today or at 
energy scales much below what was thought before \cite{extradims}. 
The reason why these 
extra dimensions could remain hidden is  that matter fields 
would be confined to a three-dimenional brane (our  visible `space')  
 whereas  gravitational fields would  live
everywhere and thus also in the extra-dimensions.

Even non-compact extra-dimensions can be envisaged 
\cite{old}, like 
in the five-dimensional model of Randall and Sundrum \cite{rs99b}, where the 
bulk is 
endowed with a negative cosmological constant. 
Remarkably, in this model, the gravity felt by observers on 
the brane will behave, at least at first approximation, like usual 
four-dimensional gravity, because of the existence of zero-mode gravitons 
effectively confined in the brane.

When one considers a brane universe in the context of cosmology, it appears 
that one does not recover the standard Friedmann equations easily 
\cite{bdl99}. 
The reason is that the matter content of the brane  enters quadratically 
in the equations governing the dynamics of the brane geometry. 
However, when one generalizes the idea of Randall and Sundrum to cosmology, i.e.
when one assumes the existence of a negative cosmological constant in the 
bulk, adjusted so as to compensate the tension of the brane, one ends up with 
a cosmological evolution that is conventional at sufficiently low energies
(\cite{cosmors} and \cite{bdel99}).

The question now is whether this conventional behaviour 
will still be recovered in brane cosmology, when one relaxes the 
assumption of homogeneity and isotropy in the brane, in other 
words when one considers {\it inhomogeneous brane cosmology}.
Up to now, almost all works (see e.g. \cite{branecosmo}) 
on the cosmological aspects of the brane scenario  have dealt 
with a homogeneous and isotropic brane universe. 
However, it is well known that 
a lot of information is contained in the cosmological perturbations of our 
Universe, and it is therefore of the uppermost importance to analyse the 
behaviour of cosmological perturbations in the context of brane cosmology.
It would be crucial to test whether brane cosmology is compatible with 
current observations of cosmological perturbations, either 
the  anisotropies of the Cosmic Microwave Background or the 
large scale structure data. It would also be  interesting to check if 
the usual mechanism of generation of cosmological perturbations via 
amplification of quantum fluctuations during an inflationary phase would
still be valid in brane cosmology (see \cite{mwbh99}).

Another motivation to study the perturbations in the cosmological context is 
the question of the stabilisation of the radion and its implication on the 
cosmological evolution in the brane \cite{cgrt99}. As shown by 
\cite{cgr99} in a non-cosmological context, a perturbative approach 
may be necessary for a full  understanding  of the radion. 

The purpose of the present paper is to develop a formalism 
that can describe the evolution of the cosmological perturbations 
in the context of brane cosmology. In order to do so, we will proceed in 
several steps. First, one must compute the perturbations of the {\it bulk} 
Einstein's equations, which relate the perturbations of the metric to the 
perturbations of the bulk matter, if there is any. For this first step, we have 
 computed the perturbed Einstein's equations in the more general 
case of a  warped spacetime with any number of dimensions. A similar 
calculation has been done recently \cite{Mukohyama00} in the case
of maximally symmetric spacetimes.

The   results valid for any dimension can  be applied to the model we wish 
to focus on: a five-dimensional spacetime. It is then useful to distinguish 
several types of  cosmological modes, following the familiar decomposition 
of  the standard theory of cosmological perturbations (see e.g. \cite{mfb}
or \cite{ks}).

At this stage, however,  the perturbations of the {\it matter 
in the brane} have not yet been taken into account. They play a r\^ole 
{\it only in the junctions conditions}: 
the jump in the derivatives (with respect to the 
fifth dimension) of the bulk metric perturbations is indeed governed 
by the matter perturbations in the brane. This is in this very unusual way that 
matter perturbations in our apparent Universe, i.e. the three-brane, 
are connected with the geometry perturbations in our Universe, i.e. simply 
the particular value on the brane of the bulk metric perturbations.
It is nevertheless interesting to notice that the present problem has 
some analogy with the question of the interaction between a domain wall 
and gravitational waves in a four-dimensional spacetime \cite{ii97}. 

The plan of the paper will be the following. In Section 2, we obtain 
the perturbations of Einstein's equation in the bulk, in the case of a general 
warped spacetime. In Section 3, we consider the five-dimensional model and 
recall what is known about the homogeneous and isotropic solutions in 
an empty bulk or with a negative cosmological constant. Section 4 is devoted 
to the bulk perturbations of the five-dimensional spacetime, whereas 
section 5 deals with the junction conditions that take into account 
the matter in the brane. Finally, we conclude in the last section.

\section{Bulk perturbations in general warped spacetime}

Although we will be ultimately interested in the perturbations of 
a five-dimensional spacetime, it is instructive to begin with a more general 
situation and to compute the perturbations of the Ricci tensor for a 
D-dimensional spacetime that can be considered as a warped product 
of a p-dimensional spacetime with a d-dimensional space.
The metric has thus the particular form 
\beq
\bar g_{AB}dx^A dx^B=\g_{ab} dx^adx^b+a^2\{x^c\} \s_{ij} dx^i dx^j,
\label{genmetric}
\eeq
where $A, B,...$  denote  global spacetime indices, $a, b,...$ 
indices of the  p-dimensional spacetime with metric $\g_{ab}$ 
,  and $i, j,...$  indices 
of the d-dimensional space with metric $\s_{ij}$.

Let us  denote $D_A$ the global covariant derivative associated with  the 
metric $\bar g_{AB}$, whereas $\D_a$ will stand for 
the covariant derivative associated with the 
metric $\g_{ab}$ and $\d_i$ for the covariant derivative associated with the 
metric  $\s_{ij}$. The nonvanishing mixed Christoffel symbols (those 
with indices of both types $a$ and $i$) are
\beq
\Gamma^i_{aj}={\partial_a a\over a}\s^i_j, \quad
\Gamma^a_{ij}=- (a\partial^a a)\s_{ij}.
\eeq
All tensors components will be decomposed in several sets, depending on the 
number of p-indices and of d-indices. And one can then decompose the 
action of the global spacetime covariant derivative $D_A$  into the action 
of the covariant derivative $\D_a$ and that of the covariant derivative $\d_i$.
For illustration, on a vector $V^A=(V^a,V^i)$, the action of the covariant 
derivative $D_A$ gives
\begin{eqnarray}
D_aV^b&=&\D_a V^b, \cr
D_a V^i&=&{1\over a} \D_a(aV^i), \cr 
D_iV^a&=&\d_i V^a-(a\partial^a a)\s_{ij} V^j, \cr
D_i V^j&=&\d_i V^j+{\partial_b a\over a}\s^i_jV^b.
\end{eqnarray}
This type of formulas can be generalized to any type of tensor by using 
the mixed Christoffel symbols given above.

Let us now consider a linear perturbation of the spacetime metric, so that 
the total metric reads
\beq
ds^2= (\bar g_{AB}+h_{AB})dx^A dx^B\equiv g_{AB} dx^A dx^B.
\label{pertmetric}
\eeq
Our purpose will now be to compute  the 
linear perturbation of the Ricci tensor, $\dd R_{AB}$.
Quite generally, the expression of  $\dd R_{AB}$ in terms 
of the metric perturbations is found to be of the following form (see e.g. 
\cite{wald})
\beq
\dd R_{AB}=-{1\over 2}D_AD_Bh-{1\over 2}D^CD_C h_{AB}+D^CD_{(B}h_{A)C},
\eeq
where $h$ denotes the trace of the perturbation $h_{AB}$, namely
\beq
h\equiv {\bar g}^{AB} h_{AB}.
\eeq

After some tedious but straightforward calculations, it is possible to express
the various components of the linearized Ricci tensor more 
specifically in terms of the components $h_{ab}$, $h_{ai}$ and $h_{ij}$, 
and of the covariant derivatives $\D_a$ and $\d_i$.
One finds
\begin{eqnarray}
\dd R_{ab}&=&-{1\over 2} \D_a\D_b h-{1\over 2}\D^c\D_ch_{ab}
+\D^c\D_{(a}h_{b)c} 
-{1\over 2}a^{-2}\d^2 h_{ab}+a^{-2} \d^k\D_{(a}h_{b)k} \cr
&&
+{d\over 2}{\D^c a\over a}\left(\D_a h_{bc}+\D_b h_{ac}-\D_c h_{ab}\right)
-a^{-1}\D_{(a}a\D_{b)}\hh,  \label{ricciab}
\\
\dd R_{ai}&=&-{1\over 2} \D_a\d_i h-{1\over 2}\D^b\D_b h_{ai}
+{1\over 2}\D^b\D_a h_{bi}+{1\over 2}\d_i\D^b h_{ab} 
+{1\over 2}a^{-2}\D_a\d^k h_{ik}-{1\over 2}a^{-2}\d^k\d_k h_{ai}\cr
&&
+{1\over 2}a^{-2}\d^k\d_i h_{ak} 
+{1\over 2}{\D_a a\over a}\d_i(h-\hat h)-{\D_a a\over a}\D^b h_{bi}
+{d\over 2}{\D^b a\over a}\D_a h_{bi}+\left(1-{d\over 2}\right)
{\D^b a\over a}\D_b h_{ai}\cr
&&
-\left(1-{d\over 2}\right)
{\D^b a\over a}\d_i h_{ab}
-{\D_a\over a^3}\d^k h_{ik}-(d-1){\D^b a\D_a a\over a^2}h_{bi}
-{\D^b\D_a a\over a} h_{bi} 
\\
\dd R_{ij}&=&-{1\over 2}\d_i\d_j h-{1\over 2}\left(a\D^a a \right)\s_{ij}
\D_a h-{1\over 2}a^{-2}\d^k\d_k h_{ij}+{1\over 2}a^{-2}\d^k\d_i h_{jk}
+{1\over 2}a^{-2}\d^k\d_j h_{ik}\cr
&&
-{1\over 2}\D^a\D_a h_{ij}+{1\over 2}\D^a\d_i h_{aj}
+{1\over 2}\D^a\d_j h_{ai}+{\D^a a\over a}\s_{ij}\d^k h_{ak}
+\left(a\D^b a\right)\s_{ij}\D^a h_{ab}\cr
&&
+\left(2-{d\over 2}\right)
{\D^a a\over a}\D_a h_{ij}+\left({d\over 2}-1\right){\D^b a\over a}\d_i
h_{bj} +\left({d\over 2}-1\right){\D^b a\over a}\d_j
h_{bi} \cr
&&
+(d-1)(\D^a a \D^b a)\s_{ij} h_{ab}+a(\D^a\D^b a)\s_{ij} h_{ab}
-2{\D^b a\D_b a\over a^2}h_{ij}, 
\label{ricciij}
\end{eqnarray}
with $\hh\equiv a^{-2}\s^{ij}h_{ij}$.

\bigskip 
{\it Gauge transformations.}

It is important to notice that perturbations which differ quantitatively 
can in fact describe the same geometry simply because they correspond 
to different systems of coordinates. Gauge transformations, 
corresponding to infinitesimal changes of coordinates 
\beq
x^A \rightarrow x^A+\xi^A,
\eeq 
 induce the following transformations for the metric perturbations,
\beq
h_{AB}\rightarrow h_{AB}+D_A\xi_B+D_B\xi_A.
\eeq
This general expression gives for the three types of metric components 
or, decomposing into the two subsystems of coordinates,
\begin{eqnarray}
h_{ab} &\rightarrow & h_{ab}+\g_{bc}\D_a\xi^c+\g_{ac}\D_b\xi^c, \cr
h_{ij} &\rightarrow & h_{ij}+a^2\left(\s_{jk}\d_i\xi^k+\s_{ik}\d_j\xi^k\right)
+2a (\partial_a a)\s_{ij}\xi^a, \cr
h_{ia}  &\rightarrow & h_{ia}+\g_{ab}\d_i\xi^b
+a^2 \s_{ij}\D_a\xi^j.
\label{gencoordtransf}
\end{eqnarray}

\section{Five-dimensional background spacetime}
In this section, we recall briefly the cosmological solutions that have 
been found in the case of five-dimensional spacetimes, with a metric 
of the form
\begin{equation}
ds^{2}=-n^{2}(\tau,y) d\tau^{2}+a^{2}(\tau,y)\gamma_{ij}dx^{i}dx^{j}
+b^{2}(\tau,y)dy^{2},
\label{metric}
\end{equation}
where the spatial three-surfaces, defined by $\tau$ and $y$ constant, are
homogeneous and isotropic and 
 $\gamma_{ij}$ is a 
 maximally symmetric 3-dimensional metric ($k=-1,0,1$ will 
parametrize the spatial curvature).

The five-dimensional Einstein equations take the usual form 
  \beq
{ G}_{\A\B}\equiv{R}_{\A\B}-{1\over 2}{R}{g}_{\A\B}
  =\kappa^2 {T}_{\A\B}, \label{einstein}
\eeq
where $T_{AB}$ is the five-dimensional energy-momentum tensor.

With the above metric, the non-vanishing 
components of the Einstein tensor 
${ G}_{\A\B}$ are found to be 
\begin{eqnarray}
{ G}_{00} &=& 3\left\{ \fda \left( \fda+ \fdb \right) - \frac{n^2}{b^2} 
\left(\fppa + \fpa \left( \fpa - \fpb \right) \right) + k \frac{n^2}{a^2} \right\}, 
\label{ein00} \\
 { G}_{\ii\jj} &=& 
\frac{a^2}{b^2} \gamma_{ij}\left\{\fpa
\left(\fpa+2\fpn\right)-\fpb\left(\fpn+2\fpa\right)
+2\fppa+\fppn\right\} 
\nonumber \\
& &+\frac{a^2}{n^2} \gamma_{ij} \left\{ \fda \left(-\fda+2\fdn\right)-2\fdda
+ \fdb \left(-2\fda + \fdn \right) - \fddb \right\} -k \gamma_{ij},
\label{einij} \\
{G}_{05} &=&  3\left(\fpn \fda + \fpa \fdb - \frac{\dot{a}^{\prime}}{a}
 \right),
\label{ein05} \\
{G}_{55} &=& 3\left\{ \fpa \left(\fpa+\fpn \right) - \frac{b^2}{n^2} 
\left(\fda \left(\fda-\fdn \right) + \fdda\right) - k \frac{b^2}{a^2}\right\},
\label{ein55} 
\end{eqnarray} 
where a prime stands for a derivative with respect to
 $y$, and a dot for a derivative with respect to $\tau$. 
The stress-energy-momentum tensor can be decomposed into two parts, 
\begin{equation}
{T}^\A_{\quad \B}  =  \check{T}^\A_{\quad \B}|_{_{\rm bulk}} 
+ T^\A_{\quad \B}|_{_{\rm brane}},
\end{equation}
where $\check{T}^\A_{\quad \B}|_{_{\rm bulk}}$ is the energy momentum tensor 
of the bulk matter and 
$T^\A_{\quad \B}|_{_{\rm brane}}$ corresponds to the matter content
in the 
brane $(y=0)$. Since we consider in this section
 only strictly homogeneous and isotropic 
geometries inside the brane, 
the latter will be necessary of  the form
\begin{equation}
T^\A_{\quad \B}|_{_{\rm brane}}= \frac{\delta (y)}{b} \mbox{diag} 
\left(-\rho,p,p,p,0 \right), 
\label{source}
\end{equation}
where the energy density $\rho$ and pressure $p$  are functions only of time.

In the case of a general stress-energy tensor for the matter on the brane, 
and assuming a time-independent metric along the fifth dimension,
 explicit solutions for the whole metric have been found in two simple cases:
the case where the bulk is empty \cite{bdl99} and the case 
where the bulk contains  a cosmological constant \cite{bdel99}, i.e.  
\begin{equation}
\check{T}^\A_{\quad \B}|_{_{\rm bulk}}= \mbox{diag} 
\left(-\rho_B,-\rho_B,-\rho_B,-\rho_B,-\rho_B \right), 
\label{bulksour}
\end{equation}
with $\rho_B=const$.

In the case of a negative bulk cosmological constant, the most realistic,
the explicit expressions for the metric components (with $b=1$ and $n(t,y=0)=1$)
are given by \cite{bdel99}
\begin{eqnarray}
a(t,y)&=&\left\{{1\over 2}\left(1+{\K52 \rho^2\over 6\rho_B}\right) a_0^2 
+{3\C  \over \K52\rho_B a_0^2} 
+ \left[ {1\over 2}\left(1-{\K52 \rho^2\over 6\rho_B}\right) a_0^2 
-{3\C  \over \K52\rho_B a_0^2}\right]\cosh (\mu y) 
\right.
\cr &&
\left. 
-{\kappa \rho\over\sqrt{ -6\rho_B}}a_0^2 \sinh(\mu |y|)\right\}^{1/2},
\label{abckgd}
\end{eqnarray}
with $
\mu=\sqrt{-{2 \K52 \over 3}\rho_B}, 
$
and
\beq
n(t,y)={\dot a(t,y)\over \dot a_0(t)}, \label{nbckgd}
\eeq
where $a_0$ is the scale factor in the brane (i.e. in $y=0$).
The two functions $a_0(t)$ and $\rho(t)$, which appear in the 
above solution,  are obtained by solving 
\beq
{\dot a_0^2\over a_0^2}={ \K52 \over 6}\rho_B+{ \kappa^4 \over 36}\rho^2
+{\C\over a_0^4}
- {k \over a_0^2},
 \label{friedfried}
\eeq
where $\C$ is a constant (of integration), and 
\beq
\dot \rho+3{\dot a_0\over a_0}(\rho+p)=0. \label{cons}
\eeq
The last equation corresponds to the ordinary energy conservation law. 
But (\ref{friedfried}), analogous to the first Friedman equation, does not 
give the usual cosmological evolution because of the quadratic term 
$\rho^2$. It is however possible to recover the standard cosmological 
evolution, at least at sufficiently late times, if one decomposes $\rho$ into 
a tension $\sigma$ and an energy density of ordinary matter living in the 
brane and if one assumes a negative cosmological constant in the bulk 
that will compensate the $\sigma^2$ term (see \cite{bdel99}).

\section{Perturbations of a five-dimensional spacetime}
After having described our reference spacetime in the previous section, 
we will now study linear perturbations about it. 
For this task,  the calculations of Section 2,  specialized to the 
case of five-dimensional spacetime, are going 
to be very useful. As it is clear from the previous 
section, the metric $\g_{ab}$ will be defined by 
\beq
\g_{ab}dx^a dx^b=-n(t,y)^2dt^2+b(t,y)^2dy^2.
\eeq
The second metric $\s_{ij}$ is simply the metric covering the three 
ordinary spatial dimensions. In a cosmological context, there are three
choices for $\s_{ij}$ depending on the spatial curvature of spacetime.
For simplicity, it will be assumed that our Universe is spatially flat, 
which means that
\beq
\s_{ij}=\delta_{ij}.
\eeq

 From now on, we also choose  to work 
 in a Gaussian normal (GN) system of coordinates 
adapted to the three-brane, in which the brane is localized in 
$y=0$ and the metric has the form
\beq
ds^2=g_{\mu\nu}dx^\mu dx^\nu +dy^2,
\eeq
where the greek indices refer to ordinary spacetime coordinates, i.e. 
the time coordinate $\tau$ and the three ordinary spatial dimensions 
$x^i$. For the background, this choice of gauge 
corresponds simply to set
\beq
b(t,y)=1.
\eeq
In the case of the perturbed spacetime, described by the metric 
(\ref{pertmetric}), the choice of  a GN system of coordinates 
will impose 
\beq 
h_{55}=h_{5\mu}=0. 
\eeq

Taking  this into account, the linearized metric (\ref{pertmetric}) 
can now be written,
quite generally, in the form
\beq
ds^2=-n^2(1+2A)dt^2+2 B_i dx^i dt 
+a^2\left(\delta_{ij}+\hat h_{ij}\right)dx^i dx^j+dy^2.
\eeq
Following Bardeen\cite{bardeen} (see also \cite{mfb}), the linearized 
quantities specified above can 
be decomposed further,  into so-called scalar, vector and tensor quantities, 
according to the expressions,
\beq
B_i=\partial_i B+\bar B_i,
\eeq
where $\bar B_i$ satisfies $\partial_i \bar B^i=0$, and
\beq
\hat h_{ij}=2C \delta_{ij}+2 \partial_i\partial_j E+2 \partial_{(i}E_{j)}
+ E_{ij},
\eeq
where $ E_{ij}$ is transverse traceless, i.e. 
$\partial_i E^{ij}=0$, $ E^i_i=0$.
The quantities $A$, $C$, $B$ and $E$ are usually refered to as 
 `scalar' perturbations, 
$\bar B_i$ and $E_i$ as `vector' perturbations, and $E_{ij}$ 
as `tensor' perturbations.

\subsection{Gauge transformation}
As  explained earlier, the perturbations defined above can be 
quantitatively different but describe the same geometry if they are
written in different coordinate systems. In order to distinguish 
gauge effects and physical degrees of freedom, it is useful 
to write down the effect of a coordinate change (or gauge
transformation) on all the metric perturbations defined above. 
These transformations follow directly from the general expressions
given in (\ref{gencoordtransf}).

We will parametrize the infinitesimal coordinate transformation by the 
vector $\xi^A=(\xi^0,\xi^i,\xi^5)$. Let us first consider the components 
$h_{55}$, $h_{05}$ and $h_{i5}$, which vanish in a GN coordinate 
system. They transform according to the laws,
\begin{eqnarray}
h_{55}&\rightarrow& h_{55}+2{\xi^5}', \cr
h_{05}&\rightarrow& h_{05}+\dot \xi^5-n^2{\xi^0}', \cr
h_{i5}&\rightarrow& h_{i5}+\partial_i\xi^5+ a^2\dd_{ij}{\xi^j}'.
\end{eqnarray}
In order to bring any coordinate system into a GN coordinate system, 
it is clear from the above relations that once one has used $\xi^5$ to 
adjust the position of the brane at $y=0$, appropriate  choices 
of $\xi^0$ and of $\xi^i$ will be required in order 
to make $h_{05}$ and $h_{i5}$ vanish. 
Note however that the GN gauge adapted to the brane is not completely 
fixed: there is  some residual gauge freedom associated with parameters 
 $\xi^0$ and $\xi^i$ that depend only on the four ordinary spacetime 
coordinates. This residual gauge freedom can be interpreted 
as possible  redefinitions of the coodinates inside 
the brane worldsheet.

Let us now consider the transformations for the other components of the 
metric perturbations. To do that, it is convenient to 
decompose the spatial vector $x^i$ into 
\beq
\xi^i=\partial^i\xi+\vxi^i
\eeq
(where $\partial^i=\dd^{ij}\partial_j$), such that $\vxi^i$ is transverse, 
i.e. $\partial_i\vxi^i=0$. 
With this decomposition, one can see that  the three scalar parameters
$\xi^0$, $\xi$ and $\xi^5$ will induce transformations in the subset 
of scalar perturbations, whereas $\vxi^i$ will act in the `vector' 
subspace.
The tensor perturbations $\bar E_{ij}$ will be left untouched by the 
gauge transformations. Specializing the expressions
(\ref{gencoordtransf}) to our particular case here, one finds the 
following transformations:

\subsubsection{Scalar gauge transformations}
\begin{eqnarray}
A&\rightarrow& A+\dot \xi^0+{\dot n\over n}\xi^0+{n'\over n}\xi^5,\cr
B&\rightarrow& B-n^2\xi^0+a^2\dot \xi, \cr
C&\rightarrow& C+{\dot a\over a}\xi^0+{a'\over a}\xi^5, \cr
E&\rightarrow& E+\xi, 
\label{scalartransf}
\end{eqnarray}
\subsubsection{Vector gauge transformations}
\begin{eqnarray}
\vB_i&\rightarrow&\vB_i+a^2\dot \vxi_i,\cr
E_i&\rightarrow&E_i+\vxi_i.
\label{vectortransf}
\end{eqnarray}

In the above transformations, we have considered the most general gauge 
transformation. But if one considers only gauge transformations 
inside the subset of GN coordinate systems, which will be the case in the 
following, then $\xi^5$ disappears from 
(\ref{scalartransf}) and the parameters $\xi^0$, $\xi$
and $\bar\xi$ cannot depend on the fifth coordinate.

Finally let us remark that one can use the remaining gauge freedom 
within the GN subset to impose some additional gauge conditions.
For example, for scalar quantities, a choice which is familiar in the 
standard theory of cosmological perturbations is to impose 
\beq
B=0, \quad E=0, 
\eeq
which corresponds to the so-called longitudinal (or Newtonian)
gauge.  However, these conditions can be imposed only on one hypersurface, 
$y=0$ say, but not everywhere in the bulk, because $B$ and $E$ a priori 
depend on the coordinate $y$.

\subsection{Perturbed Einstein equations in the bulk}

Let us now compute the equations governing the perturbations in the bulk, 
ignoring for the moment the presence of the brane.
The five dimensional Einstein's equations (\ref{einstein}), in the bulk, can be 
rewritten 
\beq
R_{AB}=\kappa^2\left(\check T_{AB}-{1\over 3}\check T g_{AB}\right).
\eeq
Therefore, the perturbed Einstein equations in the bulk have the 
form 
\beq
\dd R_{AB}=\kappa^2\left(\dd\check T_{AB}- {1\over 3}\check T h_{AB} 
-{1\over 3} g_{AB} 
\dd\check T\right).
\eeq

Since the background solutions have been found explicity, 
as recalled in Section 3, the cases of an empty bulk or 
of a cosmological constant are of particular interest, but of course, 
the present formalism applies to any bulk energy-momentum tensor.
In the case of an empty bulk, the perturbed Einstein's equations are simply
\beq
\dd R_{AB}=0,
\eeq
whereas in the case of a bulk with a cosmological constant $\Lambda=\kappa^2
\rho_B$, the perturbed Einstein's equations read 
\beq
\dd R_{AB}={2\over 3}\Lambda h_{AB}.
\eeq

The remaining task consists in computing  explicitly the components of the 
perturbed Ricci tensor. Using the expressions (\ref{ricciab}-\ref{ricciij}), 
one obtains the following expressions, which can conveniently be 
 separated in scalar, vector and tensor parts.

\subsubsection{Scalar components}
\begin{eqnarray}
\dd R_{00}^S&=&n^2\left[A''+\left(3{a'\over a}+2{n'\over n}\right) A'
+ 3 {n'\over n} C' +\left(6{a'n'\over an}+2{n''\over n}\right)A\right] \cr
&&
-3 \ddot C+3{\dot a\over a}\dot A+\left(3 {\dot n\over n}-6 {\dot a\over a}
\right)\dot C+{n^2\over a^2}\Delta A \cr
&& 
-\Delta \ddot E+\left({\dot n\over n}-2{\dot a\over a}\right)\Delta\dot E
+nn'\Delta E'+a^{-2}\Delta \dot B-{\dot n\over a^2 n}\Delta B,
\\
\dd R_{0i}^S&=& \partial_i\left\{-2\dot C+2{\dot a \over a}A
-{1\over 2}B''+
{1\over 2}\left({n'\over n}-{a'\over a}\right)B'
+\left[{1\over n^2}\left({\ddot a\over a}-{\dot n \dot a\over na}
+2{\dot a^2\over a^2}\right)-2{a'n'\over an}\right]B
\right\},
\\
\dd R_{ij}^S&=&-a^2\left\{
\left( 4{\dot a^2\over a^2n^2}- 2{\dot a\dot n\over an^3}+2{\ddot a\over an^2} \right)A+ 2
\left({a'n'\over an}+2{a'^2\over a^2}+{a''\over a}+
{\dot a \dot n\over an^3}-2{\dot a^2\over a^2n^2}-{\ddot a\over a n^2}
\right) C \right.\cr
&&
+{a'\over a}A'+\left(6 {a'\over a}+{n'\over n}\right)C'
+{\dot a\over an^2}\dot A  \cr
&&
\left. +C''+\left({\dot n\over n^3}- 6 {\dot a \over an^2}\right)\dot C 
-{1\over n^2}\ddot C
+{\dot a\over n^2 a^3}\Delta B -{\dot a\over n^2 a}\Delta\dot E +
{a'\over a}\Delta E'
\right\} \delta_{ij}-\Delta C\delta_{ij}\cr
&&
+\partial_i\partial_j\left[-A -C + {a^2\over n^2}\ddot E
+\left({3a\dot a\over n^2}-{a^2\dot n\over n^3}\right)\dot E
-a^2E''-\left(3aa'+{n'a^2\over n}\right)E'  \right.\cr
&& 
\left. +2\left({a\ddot a\over n^2}+2{\dot a^2\over n^2}-{\dot n a\dot a\over
n^3}-aa''-2a'^2-{n'aa'\over n}\right)E 
-{1\over n^2}\dot B+\left({\dot n\over n^3}
-{\dot a\over an^2}\right) B
\right]
\\
\dd R_{i5}^S&=& \partial_i\left[-A'-2C'+\left({a'\over a}-{n'\over
n}\right) A
-{1\over 2n^2}\dot B'+
{a'\over an^2}\dot B+\left({\dot n\over 2n^3}-{3\dot a\over 2n^2a}\right)
B'  \right.\cr
&&
\left.
+\left({\dot a'\over an^2}+2{\dot a a'\over n^2a^2}-{\dot n a'
\over n^3 a}\right)B
 \right],
\\
\dd R_{55}^S&=& -6 {a'\over a}C'-2{n'\over n} A'- A''- 3C''
-\Delta E''-2{a'\over a}\Delta E',
 \\
\dd R_{05}^S&=& 3\left[ {n'\over n}\dot C- \dot C' 
+{\dot a\over a}A'-{\dot a\over a}C'-{a'\over a}\dot C\right]
+{1\over 2}a^{-2}\Delta B'-{n'\over a^2 n}\Delta B \cr
&&
-\Delta\dot E' +\left({n'\over n}-{a'\over a}\right)\Delta\dot E
-{\dot a\over a}\Delta E'.
\end{eqnarray}

\subsubsection{Vector components}
\begin{eqnarray}
\dd R_{0i}^V&=&{1\over 2}\Delta \dot E_i-{1\over 2}\vB_i''+
{1\over 2}\left({n'\over n}-{a'\over a}\right)\vB_i'
+\left[{1\over n^2}\left({\ddot a\over a}-{\dot n \dot a\over na}
+2{\dot a^2\over a^2}\right)-2{a'n'\over an}\right]\vB_i-{1\over 2a^2}
\Delta \vB_i,
\\
\dd R_{i5}^V&=& {1\over 2}\Delta E_i'-{1\over 2n^2}\dot \vB_i'+
{a'\over an^2}\dot \vB_i+\left({\dot n\over 2n^3}-{3\dot a\over 2n^2a}\right)
\vB_i'+\left({\dot a'\over an^2}+2{\dot a a'\over n^2a^2}-{\dot n a'
\over n^3 a}\right)\vB_i
\\
\dd R_{ij}^V&=&{a^2\over n^2} \partial_{(i}\ddot E_{j)}
+\left({3a\dot a\over n^2}-{a^2\dot n\over
n^3}\right)\partial_{(i}\dot E_{j)}
-a^2\partial_{(i}E_{j)}'' -\left(3aa'+{n'a^2\over n}\right)\partial_{(i}E_{j)}'
\cr
&& 
+2\left({a\ddot a\over n^2}+2{\dot a^2\over n^2}-{\dot n a\dot a\over
n^3}-aa''-2a'^2-{n'aa'\over n}\right)\partial_{(i}E_{j)}\cr
&& -{1\over n^2}\partial_{(i}\dot \vB_{j)}+\left({\dot n\over n^3}
-{\dot a\over an^2}\right)
\partial_{(i} \vB_{j)}
\end{eqnarray}

\subsubsection{Tensor components}
\begin{eqnarray}
\dd R_{ij}^T&=&-{1\over 2}\Delta E_{ij}+{a^2\over 2n^2}\ddot E_{ij}
+\left({3a\dot a\over 2n^2}-{a^2\dot n\over 2n^3}\right)\dot E_{ij}
-{1\over 2}a^2E_{ij}''-\left({3\over 2}aa'+{n'a^2\over 2n}\right)E'_{ij}\cr
&& 
+\left({a\ddot a\over n^2}+2{\dot a^2\over n^2}-{\dot n a\dot a\over
n^3}-aa''-2a'^2-{n'aa'\over n}\right)E_{ij}.
\end{eqnarray}

\section{Junction conditions}
In the previous section, we have considered only 
the perturbations in the bulk. It is now time 
 to take into account the brane itself, and in particular 
the perturbations of the matter in the brane. The connection between 
the metric perturbations, living in the bulk, and the matter perturbations
confined to the brane is made via the junction conditions \cite{israel66}.

The energy-momentum tensor describing matter content of the brane 
will be written in  the form 
\beq
T^\mu_{\quad \nu}|_{_{\rm brane}}= \delta (y) 
S^\mu_\nu,
\eeq
and, in the background, it has the perfect fluid form 
\beq
S^\mu_\nu=(\rho+P)u^\mu u_\nu+Pg^\mu_\nu.
\eeq
The junction conditions are found to be given, in the five-dimensional 
case \cite{bdl99}, by 
\beq
[K_{\mu\nu}]=-\kappa^2\left(S_{\mu\nu}-{1\over 3}S g_{\mu\nu}\right), 
\label{junction}
\eeq
where the brackets denote the jump  across the brane, of 
the extrinsic curvature $K_{\mu\nu}$ and $S\equiv S^\mu_\mu$.
In a Gaussian normal coordinate system, the extrinsic curvature 
is given by the simple expression 
\beq
K_{\mu\nu}={1\over 2}\partial_y g_{\mu\nu}. \label{kgn}
\eeq

In the present context, it is convenient to decompose the junction 
conditions  (\ref{junction}) into a background part and a perturbed
part. 
The use 
of the background junction conditions is necessary to find 
 solutions of the Einstein equations when homogeneity and isotropy 
in the brane are assumed, such as those given in (\ref{abckgd}-\ref{nbckgd}). 
The junction conditions for the background have been given in 
\cite{bdl99}
\begin{eqnarray}
\frac{\jl a^\prime \jr}{a_0 b_0}&=&-\frac{\kappa^2}{3}\rho, \label{aarho}\\
\frac{\jl n^\prime \jr}{n_0 b_0}&=&\frac{\kappa^2}{3}  \left( 3p + 2\rho \right), 
\label{nnrho}
\end{eqnarray}
where the subscript $0$ for $a,b,n$ means that these functions are 
taken in $y=0$.

Let us now consider the junction conditions for the perturbations, 
which can be written
\beq
[\dd K_{\mu\nu}]=\kappa^2\left(-\dd S_{\mu\nu}
+{1\over 3} g_{\mu\nu}\dd S+{1\over 3}S h_{\mu\nu}\right), 
\eeq
Using the expression (\ref{kgn}) for the extrinsic curvature in the GN gauge 
as well as the $y \rightarrow -y$ symmetry, one ends up with the 
following condition,
\beq
h'_{\mu\nu|y=0^+}=\kappa^2\left(-\dd S_{\mu\nu}
+{1\over 3} g_{\mu\nu}\dd S+{1\over 3}S h_{\mu\nu}\right).
\label{pertjunction}
\eeq

It is then useful to decompose further these junction conditions by 
distinguishing space and time.
  Let us begin with the
perturbations of the energy-momentum tensor in  the brane. The 
perturbations for the unit four-velocity can be written
\beq
\dd u^\mu=\left\{-n^{-1}A, a^{-1} v^i\right\},
\eeq
where the expression for $\dd u^0$ follows automatically from the 
normalization condition satisfied by $u^\mu$. The perturbations 
of the energy-momentum tensor then have the following form:
\begin{eqnarray}
\dd S_{00}&=& n^2\dd\rho+2\rho n^2 A, \cr
\dd S_{0i}&=& -(\rho+P)n a v_i-\rho B_i,\cr
\dd S_{ij}&=& a^2\dd P \dd_{ij}+Ph_{ij}+a^2\pi_{ij},
\end{eqnarray}
with $v_i\equiv \dd_{ij}v^j$ and 
where $\pi_{ij}$ is the anisotropic stress tensor.
Once more it is possible to decompose the above expressions for the 
brane matter energy-momentum tensor into scalar, vector and tensor 
components, using 
\beq
v_i=\partial_i v+\bar v_i,
\eeq
with $\partial_i \bar v^i=0$ and 
\beq
\pi_{ij}=\left(\partial_i\partial_j-{1\over 3}\dd_{ij}\Delta\right)\pi 
+2\partial_{(i}\pi_{j)}+\bar\pi_{ij},
\eeq
with as usual $\pi_i$ transverse and $\bar\pi_{ij}$ transverse traceless.
Substituting the above expressions for the perturbed energy-momentum
tensor and using the background junction conditions
(\ref{aarho}-\ref{nnrho})
yields the following conditions for the perturbations
\begin{eqnarray}
A'_{|y=0^+}&=& {\kappa^2\over 6}\left(2\dd\rho+3\dd P\right),\\
(n^{-2} B_i)'_{|y=0^+}&=&\kappa^2(\rho+P){a_0\over n_0} v_i, \\
\hat h'_{ij|y=0^+}&=& -\kappa^2\left({1\over 3}\dd\rho\dd_{ij}
+\pi_{ij}\right).
\end{eqnarray}
These conditions can also  be decomposed, in a straightforward manner, 
into scalar, vector and tensor junction conditions.

\section{Conclusion}

In the present work, we have developed a formalism in order to deal with 
the evolution of cosmological perturbations in a brane universe. This formalism
has the advantage to introduce quantities that are similar to the usual 
treatments of cosmological perturbations in standard cosmology, thus making 
easier in the future a comparison between  brane cosmoloical perturbations 
and observable quantities, such as the large scale structure and the 
temperature anisotropies.  

However, the equations governing the perturbations in the brane scenario
are much more complicated than in standard cosmology. The main reason is that 
the equations for the metric perturbations, after a usual decomposition 
into ordinary (spatial) Fourier modes, will contain partial derivatives 
with respect to time and with respect to the fifth coordinate, in contrast 
with standard cosmology where one ends up with only ordinary differential 
equations with respect to time.
Solving these equations appears to be quite a challenge.

\bigskip 
Note: while the present work was being completed, two  works on the same subject
(\cite{kis00} and \cite{maartens}) have appeared on hep-th. 

\begin{acknowledgements}
I would like to thank P. Bin\'etruy, C. Deffayet, B. Carter, N. Deruelle and 
T. Dolezel for very interesting discussions.
\end{acknowledgements}

\end{document}